\documentclass{ws-ijmpe}

\def\beq{\begin{equation}}
\def\eeq{\end{equation}}
\def\nn{\nonumber}
\def\barr{\begin{eqnarray}}
\def\earr{\end{eqnarray}}
\def\lsim{\raise0.3ex\hbox{$\;<$\kern-0.75em\raise-1.1ex\hbox{$\sim\;$}}}
\def\gsim{\raise0.3ex\hbox{$\;>$\kern-0.75em\raise-1.1ex\hbox{$\sim\;$}}}
\def\bmat{\left( \begin{array}}
\def\emat{\end{array} \right)}
\def\t{\tilde}

\def\dub{\delta_{ub}}
\def\dubp{\delta_{ub'}}
\def\dcbp{\delta_{cb'}}
\def\v44{V_{4 \times 4}}
\def\v34{V_{3 \times 4}}

\def\a12{\theta_{12}}
\def\a13{\theta_{13}}
\def\a14{\theta_{14}}
\def\a23{\theta_{23}}
\def\a24{\theta_{24}}
\def\a34{\theta_{34}}

\begin{document}

\markboth{C. S. Kim}{Tree FCNC and Non-unitarity of CKM}

\catchline{}{}{}{}{}

\title{Tree FCNC and non-unitarity of CKM matrix
}

\author{\footnotesize C. S. Kim}

\address{Department of Physics, Yonsei University,
Seoul, 120-749, Korea\\
cskim@yonsei.ac.kr}

\author{\footnotesize Amol Dighe}

\address{Tata Inst. of Fundamental Research,
Mumbai 400005, India\\
amol@tifr.res.in}

\maketitle

\begin{history}
\received{(received date)}
\revised{(revised date)}
\end{history}

\begin{abstract}
We discuss possible signatures of the tree level FCNC, which results
from the non-unitarity of CKM matrix.
We first define the unitaity step-by-step, and possible
test of the non-unitaity through the 4-value-KM parametrization.
We, then, show how the phase angle of the unitary triangle would change
in case of the vector-like down quark model.
As another example of tree FCNC, we investigate the leptophobic $Z'$ model
and its application to the recent $B_s$ mixing measurements.
\end{abstract}

\section{Introduction}

The flavor sector of the standard model (SM) is not yet properly
understood, and in particular, the mechanism of charge-parity
($CP$) violation still eludes us. The Cabbibo-Kobayashi-Maskawa
(CKM) matrix\cite{ckm} parametrizes the mixing between the three
quark famlies, and provides the the only source of $CP$ violation
within the model. This is a very strong prediction of the SM.
While it is consistent with the observations till now,
it is important to look for any sources of
$CP$ violation beyond the SM and the observable signals they may manifest
themselves in.

Here, in particular, we investigate the possibility of a
non-unitary CKM matrix.
The unitarity of the CKM matrix in SM by virtue of the fact
that there are three quarks each of up and down type.
Though it is the most parsimonious explanation still consistent
with all the available data,
the presence of more quarks that interact significantly
with the existing ones has not been ruled out.
Most of the analyses of the quark matrix have been carried out in
the context of a $3 \times 3$ unitary CKM matrix,
the fits having been performed to the parameters that implicitly
assume the $3 \times 3$ unitarity. We currently have only a few
measurements available to test this assumption directly. However,
future measurements from the B factories and the hadronic
machines will provide us with means of testing this assumption
and/or constraining the non-unitary contributions.

The $3 \times 3$ CKM matrix can be non-unitary if it is an
{\it effective} mixing matrix, i.e. if it is a submatrix
of a larger matrix. This larger matrix may be, for instance,
a $n \times n$ unitary matrix representing the mixing between
$n$ generations, or a $n \times m$ mixing matrix that is only
partially unitary\footnote{ For a $n \times m$ matrix $V$ with
$n \neq m$, it is not possible to satisfy both $V^\dagger V =1$ and
$V V^\dagger =1$ simultaneously.} like the one that arises in models
with extra isosinglet bottom quarks.
Here we consider in detail the model with a single
``Vector-like down quark'' (VdQ)\cite{vdq},
in which only one isosinglet down quark is added to the SM.
Here the CKM matrix gets modified to a $3\times 4$ matrix
$V_{VdQ}$.
Vector-like fermions appear in many extensions of the SM, like the
low energy limit of $E_6$ GUT models\cite{Bando},
or models with extra spatial dimensions on the TeV scale\cite{ADD-RS}
as towers vector-like fermions of Kaluza-Klein excitations for
the SM quarks. The detailed phenomenological studies on the possible
FCNC effects, the violation of GIM mechanism, $\Delta \rho$ constraints
and $B \to X_s \gamma$ decays
from those towers of vector-like quaks  have been
investigated in Ref.\cite{Kim}.

\section{A Flexible Parametrization of CKM matrix via 4VKM}

The Cabibbo-Kobayashi-Maskawa (CKM)\cite{ckm} matrix  makes us possible to
explain all flavor changing weak decay processes and CP violating
phenomena up to now. Unitarity of the CKM matrix in the standard model (SM)
is a unique property that we cannot loosen. We can use any parametrization of
the CKM matrix as long as its unitarity is conserved. The original
parametrization for three generation  quark mixing is the
Kobayash-Maskawa (KM) parametrization. The standard
parametrization proposed by Chau and Keung\cite{Chau} is the
product of three complex rotation matrices which are characterized
by the three Euler angles, $\theta_{12},~\theta_{13},~\theta_{23}$
and an CP--violating phase $\delta_{13}$. More widely used one is the
Wolfenstein parametrization\cite{Wolfenstein}, which was suggested
as a simple expansion of the CKM matrix in terms of the four
parameters: $\lambda,~A,~\rho$ and $\eta$. It has been also known that
the CKM matrix for the three-generation case can be parameterized
in  terms of the moduli of four of its
elements\cite{Bjorken}. This
four-value-KM (4VKM) parametrization is rephasing invariant and
directly related to the measured quantities. In three generation
case we always need four independent parameters to define a unitary
$3 \times 3$ matrix, as explained, eg.
$\theta_{12},~\theta_{13},~\theta_{23}$ and $\delta_{13}$, or
$\lambda,~A,~\rho$ and $\eta$ or even only moduli of any four
independent elements of the matrix.

The 4VKM parametrization has several advantages over the other
parametrization. This parametrization doesn't need any specific
representations for the mixing angles as long as the CKM is unitary,
and no ambiguity over the
definition of its complex CP phase is present above all.
Secondly, the Jarlskog invariant quantity
$J_{cp}$ and non-trivial higher invariants can be reformulated as
functions of moduli and quadri-products\cite{Branco2}.
However, in the 4VKM parametrization
initial four-moduli input values should be fixed by experiments.
Once we set four moduli to specific values, remaining five moduli of
mixing elements are automatically fixed and we may lose some characteristic
effects from interplaying between the moduli.

Many groups have made global fits and numerical works on CKM
matrix elements with conventional representations which satisfy
unitarity\cite{CSKIM}. One of the
problems in these conventional parameterizations
is that they are {\it fully} and {\it completely} unitary and are
not flexible to include possible non-unitary properties resulted from unknown new
physics. Therefore, it is a complicate task to make a step-by-step test
to check the unitarity with experimental data if you use a unitary
parametrization. In the following,
we present three extended definitions for the unitarities of mixing matrix
$V$ in the order of the strength of the constraints:
\begin{itemize}
\item{\underline{Weak Unitary Conditions (WUC)}:} We define that
the mixing matrix $V$ is weak unitary if it satisfies
\begin{equation}
\sum_{\alpha} |V_{i\alpha}|^2 =
\sum_{j} |V_{j\beta}|^2 = 1
\mbox{  for all } i=u,c,t, \mbox{ and } \;\;\; \beta=d,s,b.
\label{WUC1}
\end{equation}
These constraints appear to be well satisfied experimentally for
the three generation case, and we start from this. Actually it was
pointed out that there is an apparent functional violation in the
available data:
$|V_{ud}|^2+|V_{us}|^2+|V_{ub}|^2<1$~\cite{Hocker}.
For such a case with $\sum_{\alpha} |V_{u\alpha}|^2 =a <1$, we can
easily generalize our method, and we just start with this new condition.
\item{\underline{Almost Unitary Conditions (AUC)}:} In addition to
the constraint Eq.~(\ref{WUC1}), if the following constraints are
satisfied
\begin{eqnarray}
\sum_{\alpha,i\neq j} V_{i\alpha}^* V_{j\alpha} =
\sum_{j, \alpha\neq\beta} V_{j\alpha}^* V_{j\beta} = 0
\mbox{  for some parts of }&& i,j=u,c,t, \nonumber\\
\mbox{ and } \;\;
&&\alpha,\beta=d,s,b,
\label{AUC1}
\end{eqnarray}
let us call the mixing matrix almost unitary. Some combinations,
which do not satisfy Eq. (2), may not make closed triangles, and
may have different areas even though making closed triangles. We
have no specific models in which the mixing matrix satisfies this
almost unitary conditions. Therefore, we will not consider the
case with AUC.
\item{\underline{Full Unitary Conditions (FUC)}:} This corresponds
to usual unitarity in which Eqs.~(\ref{WUC1}),~(\ref{AUC1}) are
satisfied {\it for all the indices}. All six unitarity triangles
from Eq.~(\ref{AUC1}) have the same areas.
\end{itemize}

As a next step,  we further assume that the mixing matrix $V$ satisfies
full unitary conditions.
Then we have six more constraints:
\begin{eqnarray}
\sum_{j=d,s,b} V_{ij} V^*_{kj} &=& 0, \hspace{1cm}
(i,k)=(u,c),(u,t),(c,t) ,  \nonumber  \\
\sum_{j=u,c,t} V_{ji} V^*_{jk} &=& 0, \hspace{1cm}
(i,k)=(d,s),(d,b),(s,b) .
\label{con3}
\end{eqnarray}
These constraints cannot be represented without introduction of
complex numbers analytically. If we know all the absolute values
of $V$, however, we can express necessary and sufficient
conditions for the constraints, Eqs.~(\ref{con3}), in a geometric way.
Eqs.~(\ref{con3}) give six unitarity triangles corresponding to
each six constraints, and all six triangles have equal area that is
directly related to the Jarlskog's rephasing invariant parameter
$J_{CP}$. If we take one of the constraints Eqs.~(\ref{con3}), for
example,
\[
\sum_{j=u,c,t} V_{jd} V^*_{jb} = 0,
\]
a triangle is composed of three sides with lengths
$|V_{ud}||V_{ub}|,|V_{cd}||V_{cb}|,$ and $|V_{td}||V_{tb}|$,
with a necessary condition
\begin{equation}
|V_{cd}||V_{cb}| \le |V_{ud}||V_{ub}| + |V_{td}||V_{tb}| ,
\label{pheno1}
\end{equation}
where the equality holds in CP conserving case.
For more general
argument, let us rewrite Eq.~(\ref{pheno1}) as follows:
\begin{equation}
l_2 \leq l_1 + l_3  ,  \label{COND2}
\end{equation}
where, as an example,  $l_1 = |V_{ud}||V_{ub}|,l_2 = |V_{cd}||V_{cb}|,$
and $l_3 = |V_{td}||V_{tb}|.$ After taking the square on both sides of
Eq.~(\ref{COND2}) we can rearrange the constraint equation as
follows:
\begin{equation}
f(l_1,l_2,l_3)\equiv 2 l_1^2 l_2^2+ 2 l_2^2 l_3^2 + 2 l_1^2 l_3^2
- l_1^4 - l_2^4 - l_3^4 \geq 0  ,  \label{COND3}
\end{equation}
where we denote newly introduced function  $f$ for later use.
Using the Heron's formula, the square of triangular
area can be rewritten as follows:
\begin{equation}
A^2 =s(s-l_1)(s-l_2)(s-l_3)= \frac{1}{16} f(l_1,l_2,l_3) ,
\label{AREA2}
\end{equation}
where $s=(l_1+l_2+l_3)/2$.
So the necessary condition~(\ref{COND3}) for
the complete triangle means non-negative value of $A^2$. The Jarlskog's
invariant parameter is written as follows:
\begin{equation}
J_{CP}=2A=\frac{1}{2}\sqrt{f(l_1,l_2,l_3)}.  \label{JCP}
\end{equation}

Three angles $\alpha,\beta,\gamma$ of the unitarity triangle,
which characterize CP violation,
are defined as follows:
\begin{eqnarray}
\alpha &=& Arg[-(V_{td} V^*_{tb})/(V_{ud} V^*_{ub})] , \\
\beta  &=& Arg[-(V_{cd} V^*_{cb})/(V_{td} V^*_{tb})] , \\
\gamma  &=& Arg[-(V_{ud} V^*_{ub})/(V_{cd} V^*_{cb})] .
\end{eqnarray}
The sum of those three angles, defined as the
intersections of three lines, would be always equal to 180$^0$,
even though the three lines may not be closed to make a triangle,
$i.e.$ in case that CKM matrix is not unitary at all.
We can also define these quantities from the area of the unitary triangle
and its sides:
\begin{eqnarray}
\sin\beta^{\prime} &=&
\frac{2A}{|V_{td}||V_{tb}||V_{cd}||V_{cb}|} , \\
\sin\gamma^{\prime} &=&
\frac{2A}{|V_{ud}||V_{ub}||V_{cd}||V_{cb}|} ,  \\
\alpha^{\prime} &=& \pi-\beta^{\prime}-\gamma^{\prime} ,
\end{eqnarray}
when the FUC is fully satisfied and the area of the triangles
can be defined from (\ref{AREA2}). Any
experimental data that indicates $\alpha\neq\alpha^{\prime}$ or
$\beta\neq\beta^{\prime}$ or $\gamma\neq\gamma^{\prime}$ means
that three generation quark mixing matrix $V$ is not fully
unitary.

\section{$B$ Meson Mixing Phase in Vector-like Down Quark Model}

The CKM matrix ($V_{CKM}$) is unitary, i.e. it satisfies
\beq
(V_{CKM})^\dagger V_{CKM} = V_{CKM} (V_{CKM})^\dagger = 1~~.
\label{ckm-unit}
\eeq
The unitarity relations satisfied by the CKM matrix elements
lead to three independent unitarity triangles with
equal areas. (The equality of the areas reflects that there is
only one source of $CP$ violation in the SM).
The measurements of the angles of these triangles provide
powerful tests for the unitarity of the CKM matrix.
After using the constraints in (\ref{ckm-unit})
and exploiting the freedom to change the relative phases of individual
quarks, $V_{CKM}$ can be parametrized with 3 real parameters and 1 complex
phase parameter.

The matrix $V_{VdQ}$ satisfies
\beq
V_{VdQ} (V_{VdQ})^\dagger = 1~~.
\label{vdq-unit}
\eeq
After exploiting the freedom to change the relative phases of
individual quarks, $V_{VDQ}$ can be completely described in terms of
6 real parameters
and 3 complex phase parameters. The extra complex phases
indicate the possibility of extra sources of $CP$ violation.
The limits on some of these extra parameters have already been
computed in Ref\cite{vdq}.

Here we concentrate on the constraints that
come from the mixing phases measured in the $B_d$-$\bar{B}_d$
and $B_s$-$\bar{B}_s$ systems. In the SM, these phases
correspond respectively to the angles
\beq
\beta \equiv \rm{Arg}\left(-\frac{V_{cb}^* V_{cd}}{V_{tb}^* V_{td}}
\right) ~~,~~
\chi \equiv \rm{Arg}\left(-\frac{V_{cb}^* V_{cs}}{V_{tb}^* V_{ts}}
\right)
\label{beta-chi-def}
\eeq
of the unitarity triangles. In particular, the time-dependent
$CP$ asymmetry in $B_d(t) \to J/\psi K_{S/L}$ measures $\sin(2\beta)$
and the the time-dependent $CP$ asymmetry in
$B_s(t) \to J/\psi \eta^{(')}$ or $B_s(t) \to J/\psi \phi$
would give $\sin(2\chi)$ (after taking care of the significant lifetime
difference in the $B_s$ system).
Whereas $\sin(2\beta)$ has already been measured at the B-factories\cite{psiK},
the value of $\sin(2\chi)$, predicted to be
small in the SM ($\chi \approx -0.015$), has not yet been measured.

One of the clearest signals of new physics sources of $CP$ violation
would be a value of $\chi$ that is much higher than the SM prediction.
In this paper, we shall examine if such a large value of $\chi$
is possible under the VdQ model.

\subsection{The parametrization}
\label{param}

The 6 real and 3 phase parameters that describe the quark mixing
matrix in the VdQ model can be chosen to be the six rotation
angles $\theta_{ij}, ~(1 \leq i <  j \leq 4)$ and three phases
$e^{i\delta_X}, ~ X \in \{ub, ub', cb' \}$. Here we denote the
isosinglet down quark by $b'$. The matrix $V_{VdQ}$ can be
written as
\beq
V_{VdQ} = K \cdot V_{4G}~~,
\eeq
where
\beq
K = \left( \begin{array}{cccc}
1 & 0 & 0 & 0 \\
0 & 1 & 0 & 0 \\
0 & 0 & 1 & 0
\end{array} \right)
\eeq
and $V_{4G}$ is the extension of $V_{CKM}$ for four generations:
\barr
V_{4G} & \equiv &
R_{34}(\theta_{34}) \cdot
\Phi(0,-\dcbp, 0,0) \cdot R_{24}(\theta_{24}) \cdot \Phi(-\dubp,\dcbp, 0,0)
\cdot \nonumber \\
& & \cdot R_{14}(\theta_{14}) \cdot \Phi(\dubp,0, 0,0) \cdot
R_{23}(\theta_{23}) \cdot \Phi(-\dub,0,0,0) \cdot \nonumber \\
& & \cdot R_{13}(\theta_{13}) \cdot  \Phi(0,0,\dub,0)
\cdot R_{12}(\theta_{12})~~,
\label{expansion}
\earr
Here $R_{ij}$ represents the rotation in $i-j$ plane, and
$\Phi(\delta_1,\delta_2,\delta_3,\delta_4)$ is the diagonal matrix
with diagonal elements $e^{i\delta_i}, ~1 \leq i \leq 4$.
Note that $V_{VdQ}$ is just $V_{4G}$ with its fourth row removed.
Also, putting $\theta_{14}=\theta_{24}=\theta_{34}=0$ reduce
$V_{4G}$ as well as $V_{VdQ}$ to a $3 \times 3$ submatrix $V_{CKM}$
and zeroes as the remaining elements.

Taking a clue from the hierarchy of the measured mixing angles, let
us use the sine of the Cabbibo angle, $\lambda \approx 0.22$,
as the expansion parameter. The angles already present in the
CKM matrix are
\beq
\sin(\theta_{12})\equiv \lambda ~,~ \sin(\theta_{23}) \equiv A \lambda^2
~,~ \sin(\theta_{13})\equiv A C \lambda^3 ~~.
\label{old-angles}
\eeq
Keeping with the philosophy of expanding in $\lambda$, we shall
parametrize the new physics angles as
\beq
\sin(\theta_{14})\equiv p \lambda^3 ~,~ \sin(\theta_{24}) \equiv q \lambda^2
~,~ \sin(\theta_{34})\equiv r \lambda~~.
\label{new-angles}
\eeq
This reflects the assumption that the mixing between two generations
follows a hierarchical pattern, i.e. the further apart the generations,
the smaller the mixing between them.

We expand the elements of $V_{VdQ}$ in powers of $\lambda$ such that each
element is accurate upto a multiplicative factor of
$[1 + {\cal O}(\lambda^3)]$).
The definitions of the parameters in (\ref{old-angles}) and (\ref{new-angles})
translate to
\beq
\begin{tabular}{lll}
$V_{us}  =  \lambda$ , &
$V_{cb}  = A \lambda^2$ , &
$V_{ub}  =  A \lambda^3 C e^{-i\dub}$ , \\
$V_{ub'} =  p \lambda^3 e^{-i\dubp}$ , &
$V_{cb'}  = q \lambda^2 e^{-i\dcbp}$ , &
$V_{tb'}  = r \lambda $ .
\end{tabular}
\eeq
These give rise to the other elements of $V_{VdQ}$ as:
\barr
V_{ud} & = & 1 - \frac{\lambda^2}{2} + {\cal O}(\lambda^4) \label{vud}  \\
V_{cd}  & = &  -\lambda + {\cal O}(\lambda^5)   \\
V_{cs}  & = &  1 - \frac{\lambda^2}{2}+ {\cal O}(\lambda^4) \\
V_{td} & = & A \lambda^3 \left( 1 - C e^{i\dub} \right)
+ r \lambda^4 \left( q e^{i\dcbp} - p e^{i\dubp} \right) \nonumber \\
& & + \frac{A}{2} \lambda^5 \left( -r^2 + (C + C r^2) e^{i\dub} \right)
+ {\cal O}(\lambda^6) \\
V_{ts} & = & -A \lambda^2 - q r \lambda^3 e^{i\dcbp}
+ \frac{A}{2} \lambda^4 \left( 1 + r^2 - 2 C e^{i\dub} \right)
+ {\cal O}(\lambda^5)\\
V_{tb} & = & 1 - \frac{r^2 \lambda^2}{2}+ {\cal O}(\lambda^4) \label{vtb}
\earr
Note that in the limit $p=q=r=0$, only the elements present in
$V_{CKM}$ retain nonvanishing values, and the above expansion
corresponds to the Wolfenstein parametrization\cite{Wolfenstein} with
$C= \sqrt{\rho^2 + \eta^2}$ and $\dub = \tan^{-1}(\eta/\rho)$.

The matrix $V_{VdQ}$ is also parametrized often (e.g. see Ref \cite{vdq})
in terms of the parameters (in addition to the usual CKM parameters)
\beq
\begin{tabular}{lll}
$
D^2_d = |V_{t'd}|^2~,$ &
$D^2_s = |V_{t's}|^2~$, &
$D^2_b = |V_{t'b}|^2~,$ \\
$U_{sd} = -V_{t's}^* V_{t'd}$ &
$U_{bs} = -V_{t'b}^* V_{t's}$ &
$U_{bd} = -V_{t'b}^* V_{t'd}$
\end{tabular}
\eeq
where the elements $V_{t'q}$ are given in our notation by
\barr
V_{t'd} & = & \lambda^3 \left( q e^{i\dcbp} - p e^{i\dubp} \right)
+ A r  \lambda^4 \left( 1 + C e^{i\dub} \right) \nonumber \\
& & + \frac{\lambda^5}{2} \left(p e^{i\dubp}  - q r^2 e^{i\dcbp} +
pr^2   e^{i\dubp} \right)+ {\cal O}(\lambda^6) \\
V_{t's} & = & q \lambda^2  e^{i\dcbp} + A r \lambda^3 \nonumber \\
& & + \lambda^4 \left( -p e^{i\dubp} + \frac{q}{2} e^{i\dcbp} +
\frac{q r^2}{2} e^{i\dcbp} \right)+ {\cal O}(\lambda^5) \\
V_{t'b} & = & - r \lambda+ {\cal O}(\lambda^4) \\
V_{t'b'} & = & 1 - \frac{r^2 \lambda^2}{2} + {\cal O}(\lambda^4)
\earr

We already have strong direct bounds on the magnitudes of the elements
of the CKM matrix. From Ref\cite{pdg}, we can derive
%
%
\beq
0.216 < \lambda < 0.223 ~,~
0.76 < A  < 0.90 ~,~
0.23 < C < 0.59~~.
\label{lambda-a-c}
\eeq
at 90\% C.L. from the direct measurements of $|V_{us}|, |V_{cb}|$ and
$|V_{ub}/V_{cb}|$, which do not assume the unitarity of $V_{CKM}$.
Combining the direct measurements of the magnitudes of the elements
in the first two rows with the unitarity constraints (\ref{vdq-unit}),
we get the 90\% C.L. bounds on $|V_{ub'}|$ and $|V_{cb'}|$ as
\beq
|V_{ub'}| < 0.094 ~~,~~
|V_{cb'}| < 0.147 ~~.
\label{bd-ubp}
\eeq
These correspond to $ p < 9.0~,~q < 3.05$.

In addition, a strong constraint is obtained on
the combination $X_{bb}^L \equiv (V_{CKM}^\dagger V_{CKM})_{bb}$
through the measurements involving $Z \to b\bar{b}$: we have
$X_{bb}^L = 0.996 \pm 0.005$\cite{delaguila}.
This translates to $|V_{tb'}| < 0.11 $ at 90\% C.L., which
corresponds to $r < 0.5 $.

\subsection{Mixing phases in the VdQ model}
\label{vdq-phases}

In the VdQ model, the box diagrams that contributes the phase to the
$B_d$-$\bar{B}_d$ and $B_s$-$\bar{B}_s$ mixing are the same as those
in the SM. Therefore, the mixing phases are simply
\beq
\t\beta \equiv \rm{Arg}\left(-\frac{\t V_{cb}^* \t V_{cd}}
{\t V_{tb}^* \t V_{td}}
\right) ~~,~~
\t \chi \equiv \rm{Arg}\left(-\frac{\t V_{cb}^* \t V_{cs}}
{\t V_{tb}^* \t V_{ts}}
\right) ~~.
\label{beta-chi-vdq}
\eeq
We use the superscript $~\t{}~$ to denote the quantities in
the VdQ model.

\subsubsection{Unitarity relations involving $\beta$ and $\tilde\beta$}
\label{sec-beta}

In the standard model, the unitarity (\ref{ckm-unit})
of the $3 \times 3$ CKM matrix implies unitarity triangle relations.
The unitarity constraints (\ref{vdq-unit}) on the $3 \times 4$
matrix $V_{VdQ}$ do not lead to any unitarity triangles.
However, the $3 \times 3$ unitarity triangles get modified
in a predictable and controlled manner. We demonstrate this
in this subsection and the next.

\begin{figure}[t]
\epsfig{file=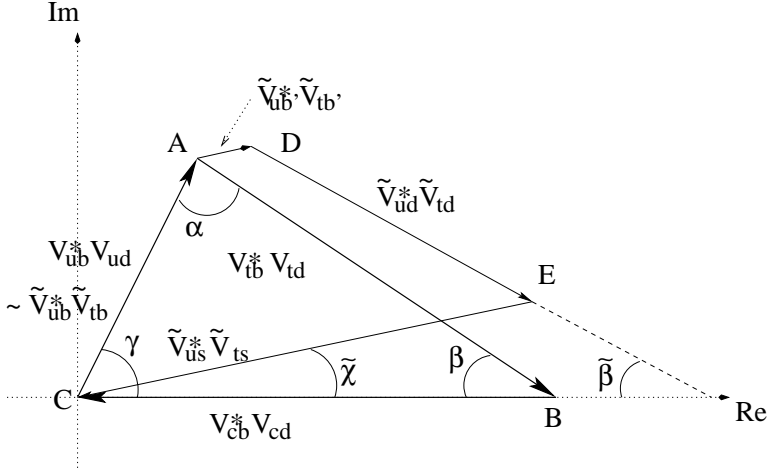,width=4in}
\caption{The unitarity triangle (ABC) in the SM with angles $\alpha, \beta,
\gamma$ and the corresponding unitarity quadrilateral (ACED) in the
VdQ model.
\label{fig-ut1}}
\end{figure}

Let us first start with the ``standard'' unitarity triangle in the SM,
which arises from the equation
\beq
V_{ub}^*V_{ud} + V_{cb}^*V_{cd} + V_{tb}^*V_{td} = 0 ~~,
\eeq
which is true in the SM. The angles of this unitarity triangle
(Fig.~\ref{fig-ut1}) are defined as
\beq
\alpha \equiv \rm{Arg}\left(-\frac{V_{tb}^* V_{td}}{V_{ub}^* V_{ud}}
\right) ~,~
\beta \equiv \rm{Arg}\left(-\frac{V_{cb}^* V_{cd}}{V_{tb}^* V_{td}}
\right) ~,~
\gamma \equiv \rm{Arg}\left(-\frac{V_{ub}^* V_{ud}}{V_{cb}^* V_{cd}}
\right) ~.
\label{abg-def}
\eeq

A related unitarity relation in the VdQ model is
\beq
\t V_{ud}^* \t V_{td} + \t V_{us}^* \t V_{ts} + \t V_{ub}^* \t V_{tb} +
\t V_{ub'}^* \t V_{tb'}  = 0 ~~,
\label{quad1}
\eeq
which may be called a ``quadrilateral'' relation. This quadrilateral
may be superimposed on the SM unitarity triangle as shown in
Fig.~\ref{fig-ut1}.

The following things should be noted. The bases of the triangle
and the quadrilateral make an angle
${\rm Arg}[(\t V_{cb}^* \t V_{cd})/(\t V_{us}^* \t V_{ts})]$
with each other, which is equal to $\t \chi$.
The  ``left'' edges of the two figures coincide since
${\rm Arg}(\t V_{ub}^* \t V_{tb})  = {\rm Arg}(V_{ub}^*V_{ud})
= \dub$.

The measurements of the $CP$ asymmetry in the decay
$B_d(t) \to J/\psi K_{S/L}$ at the $B$ factories\cite{psiK}
measure $\sin(2\beta_{\psi K}) = \sin(2 \t \beta) = 0.734 \pm 0.054$.
In Fig.~\ref{fig-ut1}, $\t \beta$ is the angle made by the ``right''
leg of the quadrilateral with the real axis.

We can estimate the systematic error in the measurement of $\beta$
due to the VdQ extension to be
\barr
\Delta \beta & \equiv & \t \beta - \beta =
{\rm Arg}\left( \frac{\t V_{ud}^* \t V_{td}}{V_{tb}^* V_{td}} \right)
= {\rm Arg}\left(\frac{\t V_{td}}{V_{td}}\right) \nn \\
& \approx & {\rm Arg}\left[ 1 + \frac{r \lambda}{A} \left(
\frac{ q e^{i\dcbp} - p e^{i\dubp}}{1 - C e^{i\dub}} \right) \right]~~.
\earr
Thus, we estimate $\Delta \beta \lsim \lambda$. Since the current
experimental error corresponds to $\Delta \beta \approx 0.04$, the
deviation due to the VdQ extension may be important.
It is, however, not possible to identify this deviation from the
measurement of $\beta_{\psi K}$ alone.

\subsubsection{Unitarity relations involving $\chi$ and $\t \chi$}
\label{sec-chi}

\begin{figure}[t]
\epsfig{file=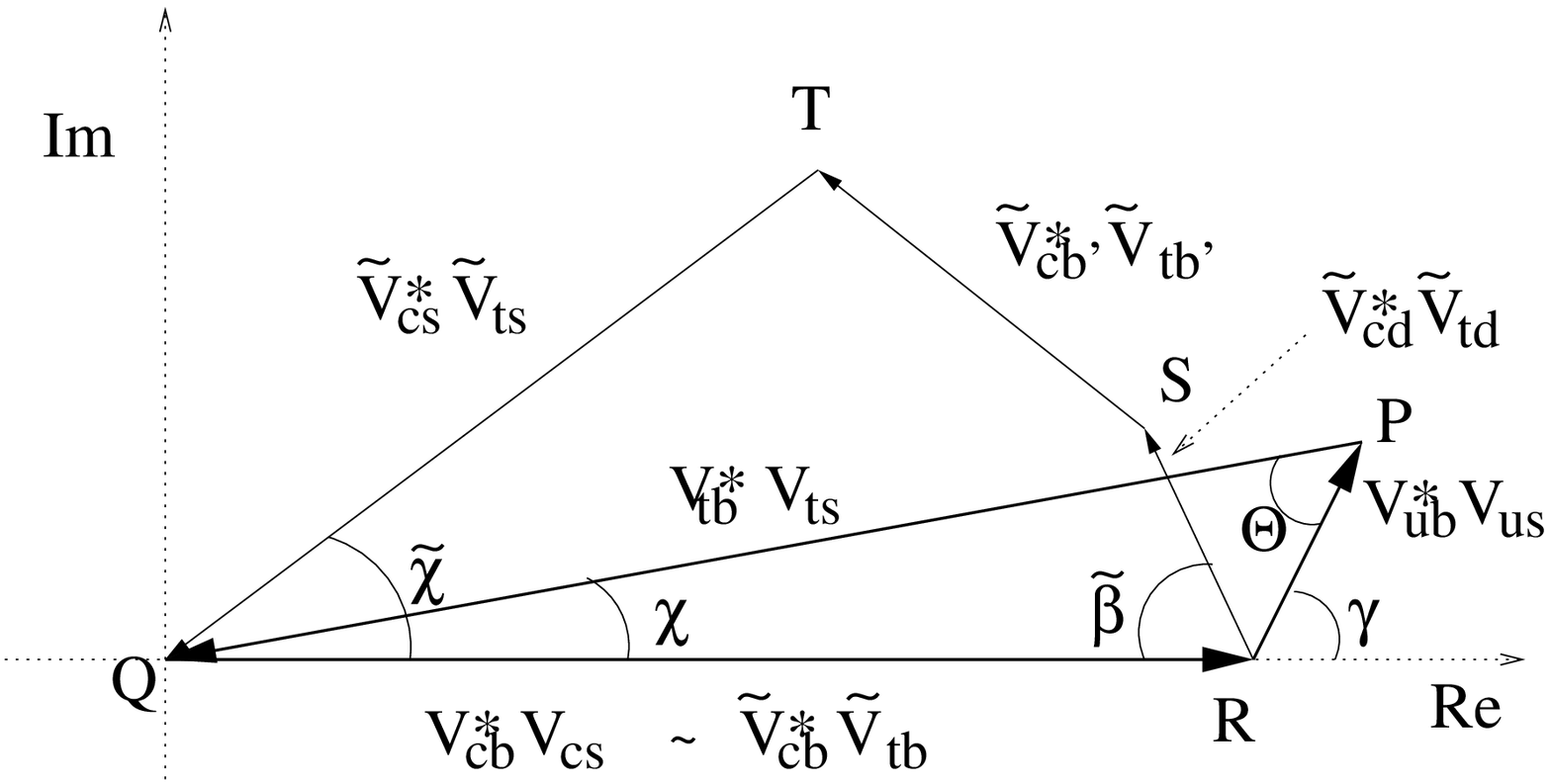,width=4in}
\caption{The ``squashed'' unitarity triangle (PQR) in the SM and the
corresponding unitarity quadrilateral (QRST) in the VdQ model.
\label{fig-ut2}}
\end{figure}

The ``squashed'' unitarity triangle in the SM arises from the equation
\beq
V_{ub}^*V_{us} + V_{cb}^*V_{cs} + V_{tb}^*V_{ts} = 0 ~~.
\eeq
The angles of this unitarity triangle (Fig.~\ref{fig-ut2}) are
\beq
\chi \equiv \rm{Arg}\left(-\frac{V_{cb}^* V_{cs}}{V_{tb}^* V_{ts}}
\right) ~,~
\Theta \equiv \rm{Arg}\left(-\frac{V_{tb}^* V_{ts}}{V_{ub}^* V_{us}}
\right) = \gamma - \chi ~,~
\pi - \Theta - \chi ~~.
\label{chidef}
\eeq
Unitarity of the CKM matrix forces the angle $\chi$ (also called in the
literature by various names like $\delta \phi, \phi_s, 2\delta\gamma,
\beta_s$) to be very small: in fact, the relation
\beq
\sin\chi \approx \left| \frac{V_{us}}{V_{ud}} \right| ^2
\frac{\sin\beta \sin(\gamma+\chi)}{\sin(\beta+\gamma)}~~
[1 + {\cal O}(\lambda^4)]
\label{sinchi}
\eeq
is a true test of unitarity (as opposed to the relation
$\alpha+\beta+\gamma = \pi$, which is trivially true by the definitions
in (\ref{abg-def})). This gives $\chi \approx 0.015$ in the SM.
If the value of $\chi$ (which also is the $B_s$-$\bar{B}_s$
mixing phase) is indeed this small, it may be hard to measure it
even at the LHC. However, as we shall see, in the
extensions of the SM that defy $3\times 3$ CKM unitarity, it is indeed
possible to get higher values of this phase.

The unitarity ``quadrilateral'' relation in the VdQ models that is
relevant here is
\beq
\t V_{cd}^* \t V_{td} + \t V_{cs}^* \t V_{ts} + \t V_{cb}^* \t V_{tb} +
\t V_{cb'}^* \t V_{tb'}  = 0 ~~,
\eeq
This quadrilateral may be superimposed on the SM unitarity triangle
as shown in Fig.~\ref{fig-ut2}.
The bases of both the triangle and the quadrilateral coincide
since ${\rm Arg}(\t V_{cb}^* \t V_{tb}) = {\rm Arg}(V_{cb}^* V_{cs})
= 0$.

The systematic error introduced in the measurement of $\chi$ due
to the VdQ extension is
\barr
\Delta \chi & \equiv & \t \chi - \chi =
{\rm Arg}\left( \frac{\t V_{cs}^* \t V_{ts}}{V_{tb}^* V_{ts}} \right)
= {\rm Arg}\left(\frac{\t V_{ts}}{V_{ts}}\right) \nn \\
& \approx & {\rm Arg}\left( 1 +  \frac{q r \lambda}{A}e^{i\dcbp} \right)
\sim \frac{q r \lambda}{A} ~~.
\earr
This, $\Delta \chi$ (or $\t \chi$) can be as large as the order of
$\lambda \approx 0.2$. In this case, the contribution due to the
VdQ extension dominates over the SM contribution by about an order
of magnitude. If the value of $\t \chi$ is indeed this high, it can
be definitely measured at the LHC or perhaps
even at the Tevatron.

At the same time, the value of $\t\chi$ is restricted to be ${\cal O}
(\lambda)$. Thus, too large a measured value of the $B_s$-$\bar{B}_s$
mixing phase will be able to rule out the VdQ model.

The two major modes employed by the $B$ factories for the measurement
of the $B_d$-$\bar{B}_d$ mixing phase are
$B_d(t) \to J/\psi K_{S/L}$ and $B_d(t) \to \phi K_{S/L}$.
The central values of the phases obtained through these two modes
currently differ a lot from each other, however these values are still
consistent, given large statistical errors on the measurement of the phase
through $\phi K_{S/L}$.

The time dependent $CP$ asymmetries in the $J/\psi K_{S/L}$ and
$\phi K_{S/L}$ actually measure different quantities. Whereas the
former measures $\sin(2\beta)$, the latter measures
$\sin(2\beta+2\chi)$.
Since $\chi$ is very small in the SM, these two measurements are
expected to be identical, but in the models where $\chi$ can take
larger values, these measurements have to be considered as
independent ones.

Thus, in the VdQ model, the measurements of the $CP$ asymmetries
in these two decays\cite{psiK} imply
\barr
\sin(2\beta_{\psi K})  &=&  \sin(2\t\beta)~~,
\label{beta-psiK} \\
\sin(2\beta_{\phi K}) &=& \sin(2\t\beta + 2\t\chi)~~.
\label{beta-phiK}
\earr
Since within the SM the difference between the asymmetries
in $B^0 \to J/\psi K_S$ and $B^0 \to \phi K_S$
is expected to be
$$
|\sin(2\beta_{\psi K})-\sin(2\beta_{\phi K})| \le \sin(2\chi)
\approx {\cal O}(\lambda^2)~~,
$$
if this value shows any significant deviation from the SM prediction, then
it may reveal new physics effects.
The difference between these two measurements in VdQ
puts bounds on the value of $\t \chi$.
Although it is consistent with zero at the moment, more data on
$\phi K_{S/L}$ will reduce the errors, opening up a way for not
only the detection of new physics, but also for the direct
measurement of a new physics quantity.
If this measurement coincides with the one through the $B_s$--$\bar{B}_s$
mixing, it will be a strong signal for the VdQ model.

We note that supposing the mode $B^0 \to \phi K_S$ reveals new physics
effects, one might expect that the other modes having the same internal
quark level process $b \to s \bar s s~(e.g.,~B^0 \to \eta' K_S)$
would reveal similar effects\cite{Kim2}. Interestingly, the
recent measurements of CP asymmetries in $B^0 \to \eta' K_S$
by  Belle\cite{phiK} agree well with the results of
$\sin(2 \beta_{\psi K})$ given in Eq (48).
However, as is well known, the strong interaction physics behind
the decay $B^0 \to \eta' K_S$ is not well understood yet due to
the uncertainties from anomalous gluonic contributions,
intrinsic charm contents, involvements of spectator quark, etc.,
so we cannot yet draw any definite conclusions concerning any new physics
effects in this decay.

\section{Electroweak Penguin and Leptophobic $Z'$ Model}

Since in the standard model (SM)
the flavor changing neutral current (FCNC) processes
appear at the quantum level
with suppression factors by small electroweak gauge coupling,
CKM matrix elements, and loop momenta,
they are very sensitive to probe new physics (NP) beyond the SM
which have an enhancement factor in the coupling
or have tree-level FCNCs.

The decay of $B$ mesons accumulated largely at asymmetric $B$-factories
and Tevatron give an opportunity to probe NP models via the rare $B$ decays
induced by FCNCs.
Recently, among several sources for FCNCs in the $B$ decays,
the electroweak (EW) penguin operators have drawn much interest.
For example, the QCD penguin dominant $B\to K\pi$ decays
appear to be very interesting
since branching ratios (BRs) and mixing-induced CP asymmetry
allow much room for large NP contribution, especially
in the EW penguin sector~\cite{Kim:2005jp,Wu:2006ur}.

Most of models contributing to
the EW penguin sector have a severe constraint from
the $b\to s\gamma$ decay.
While, models such as the $Z^\prime$ model are free from such constraints
although they predict the EW penguin contributions.
In order to probe such NP models, one must resort to nonleptonic decays
or very rare process $B\to M \nu\bar{\nu} (M=\pi,K,\rho,K^\ast)$.
However, nonleptonic decays might be inefficient since they suffer from
large hadronic uncertainties and EW penguins contributions are subdominant
in nonleptonic decays.

Recently, D{\O}~\cite{D0} and CDF~\cite{Gomez-Ceballos:2006qm} Collaborations at
Fermilab Tevatron have reported
the first observation of the mass difference $\Delta m_s$
in the $B_s^0 -\overline{B}_s^0$ system which induced by the $b\to s$ FCNC:
 \begin{eqnarray}
  \textrm{D{\O}}~&:&~\quad 17 ~ \textrm{ps}^{-1} < \Delta m_s < ~21 ~ \textrm{ps}^{-1}
                        ~~\left( 90 \% ~ \textrm{C.L.}\right) ,
  \nonumber\\
  \textrm{CDF}  ~&:&~\quad \Delta m_s = 17.33_{-0.21}^{+0.42} (\textrm{stat.})
                                    \pm 0.07 (\textrm{syst.})~ \textrm{ps}^{-1}.
 \end{eqnarray}
These measurements may give strong constraints on the NP models,
which predict $b \to s $ FCNC transitions~\cite{MSSM,RS-kim}.

In the present work, we focus on the leptophobic
$Z^\prime$ model motivated from the flipped SU(5) or string-inspired
$E_6$ models as a viable NP model. In Sec.~\ref{sec2},
we briefly introduce the leptophobic $Z^\prime$ model.
Section~\ref{sec3} deals with
$B \rightarrow M \nu \bar{\nu}~ (M= \pi, K, \rho, K^*)$ decays
within the leptophobic $Z^\prime$ model.
We investigate implications of $\Delta m_s$ measurements on this model
in Sec.~\ref{sec4}.

\subsection{Leptophobic $Z^\prime$ model and FCNC
\label{sec2}}

In many new physics scenarios containing an additional $U(1)^\prime$ gauge
 group at the low energy, the new neutral gauge boson $Z^\prime$
would have a property of leptophobia, which means that
the $Z^\prime$ boson does not couple to the ordinary SM charged leptons.
In flipped SU(5)$\times$U(1) scenario~\cite{Lopez:1996ta},
leptophobia of the $Z^\prime$ boson can be given naturally because
the neutrino is subject to the different representation with the charged
leptons.
 Another scenario for leptophobia can be found in the $E_6$ model with
kinetic mixing, where in this model leptophobia is somewhat accidental.
After breaking of the $E_6$ group,
the low energy effective theory contains an extra $\textrm{U(1)}^\prime$
which is a linear combination of $\textrm{U(1)}_\psi$ and $\textrm{U(1)}_\chi$
with a $E_6$ mixing angle $\theta$\cite{Rizzo:1998ut}.
Then, the general interaction Lagrangian of fermion fields and $Z^\prime$
gauge boson can be written as
\begin{equation}
{\cal L}_{\rm int} = - \lambda \frac{g_2}{\cos \theta_W}
\sqrt{\frac{5 \sin^2 \theta_W}{3}}
\bar{\psi} \gamma^\mu \left( Q^\prime + \sqrt{\frac{3}{5}}\delta Y_{SM} \right)
\psi Z_\mu^\prime ~,
\end{equation}
where the ratio of gauge couplings $\lambda = g_{Q^\prime}/g_Y$,
and $\delta=-\tan \chi/\lambda$\cite{Rizzo:1998ut}.
Since the general fermion-$Z^\prime$ couplings depend on two free parameters,
$\tan \theta$ and $\delta$, effectively,
the $Z^\prime$ boson can be leptophobic within an appropriate embedding of
the SM particles\cite{Rizzo:1998ut,Leroux:2001fx}.

Assuming  $V_L^d = 1$ in the $E_6$ model and  flipped SU(5) model,
only $Z^\prime$-mediating FCNCs in the right-handed down-type quarks
survive.
Then, one can get the FCNC Lagrangian for the $b\to q (q=s,d)$
transition~\cite{Jeon:2006nq}
\begin{equation}
{\cal L}_{\rm FCNC}^{Z^\prime} = - \frac{g_2}{2 \cos \theta_W}
U_{qb}^{Z^\prime} \bar{q}_R \gamma^\mu b_R Z_\mu^\prime ,
\end{equation}
where all the theoretical uncertainties including the mixing
parameters are absorbed into the coupling $U_{qb}^{Z^\prime}$. The
coupling $U_{sb}^{Z^\prime}$ has in general CP violating complex
phase, which we denote as $\phi_{sb}^{Z^\prime}$. We note that the
leptophobic $Z^\prime$ boson is not well constrained by
experiments including the charged leptons such as $b\to s \ell^+
\ell^-$ or $B_{(s)} \to \ell^+ \ell^-$, while the typical new
physics models are strongly constrained by such experiments.

\subsection{Exclusive $B\to M\nu\bar{\nu} $ Decays
\label{sec3}}

In this section, we consider the $B\to M\nu\bar{\nu} $
decays in the leptophobic $Z^\prime$ model.
The $B\to M\nu\bar{\nu}$ decays are measured via the scalar or vector meson with
the missing energy signal.

Theoretical estimates for BRs of
the $B\to M\nu\bar{\nu} $ decays
in the SM  are $0.22^{+0.27}_{-0.17}$, $5.31^{+1.11}_{-1.03}$,
$0.49^{+0.61}_{-0.38}$, and $11.15^{+3.05}_{-2.70}$ in units of $10^{-6}$,
respectively.
While experiments by the Belle and BaBar Collaborations have reported
only upper limits on BRs of
$B\to K\nu \bar{\nu}$ and $B\to \pi \nu \bar{\nu}$ decays
\cite{Abe:2005bq,Aubert:2004ws},
where the experimental bounds are about 7 times larger than the SM expectation
for the $K$ production and much larger by an order of $10^3$ for the $\pi$
production.

\begin{figure}
\begin{center}
\begin{tabular}{cc}
\psfig{file=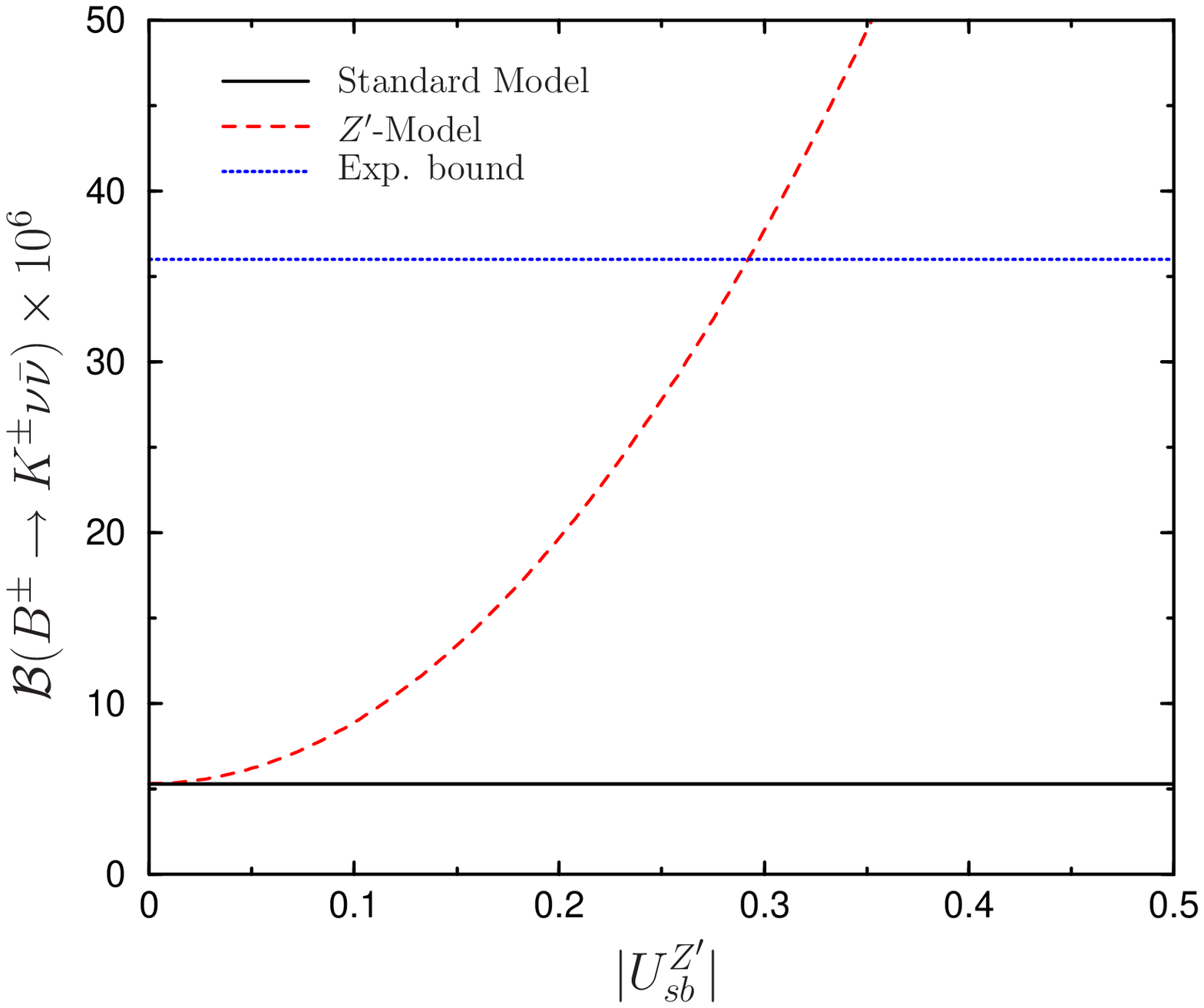,width=6cm}~~~&
\psfig{file=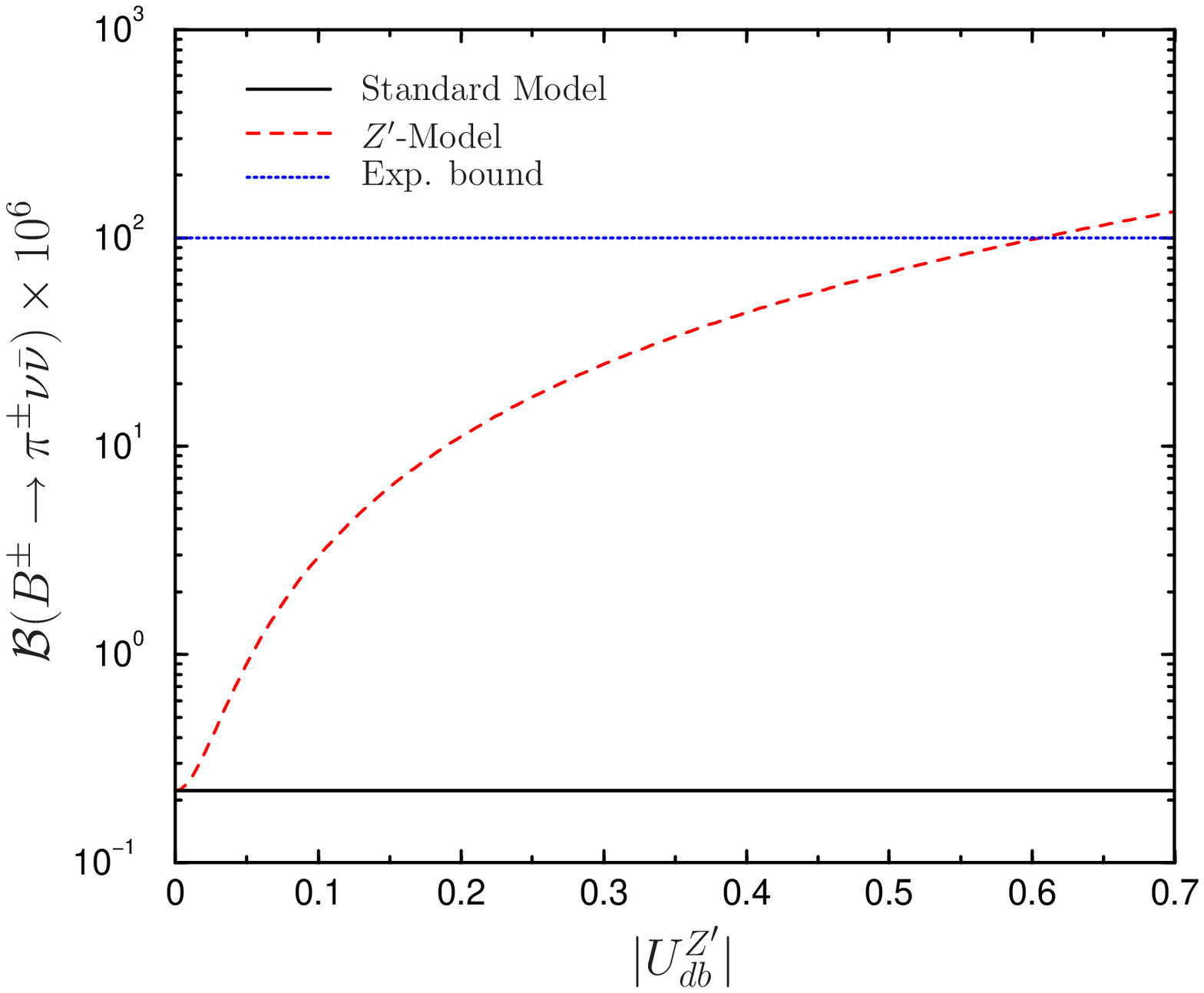,width=6cm}~~~\\[-0.0ex]
\textbf{(a)}&\textbf{(b)}
\end{tabular}
\vspace*{1pt}
\caption{ \label{fig1}
Branching ratios for (a)~$B^\pm \to K^\pm \nu\bar{\nu}$ and
(b)~$B^\pm \to \pi^\pm \nu\bar{\nu}$,
where $\nu$ can be the ordinary SM neutrinos or right-handed neutrinos.
}
\end{center}
\end{figure}

The leptophobic $Z^\prime$ model can yield same signals as
$B\to K\nu_{\rm SM} \bar{\nu}_{\rm SM}$ at detectors
via the production of a pair of right-handed neutrinos instead of
the ordinary SM neutrinos.
In Fig.~\ref{fig1}, we present our predictions for the BRs in the leptophobic
$Z^\prime$ model as a function of the effective coupling $|U_{qb}^{Z^\prime}|$,
where the mass of the $Z^\prime$ boson is assumed to be 700 GeV.
The solid and dotted lines represent the estimates in the SM and the current experimental bounds, respectively.
The dashed line denotes the expected BRs in the leptophobic $Z^\prime$ model.
In spite that we choose a specific mass for the $Z^\prime$ boson,
the present analysis can be easily translated through the corresponding changes
in the effective coupling $|U_{qb}^{Z^\prime}|$
for different $Z^\prime$ boson mass.
We extract the following constraints for the FCNC couplings from
Fig.~\ref{fig1}
\begin{equation}
|U_{sb}^{Z^\prime}| \leq 0.29 , ~~
|U_{db}^{Z^\prime}| \leq 0.61 ,
\label{Ubound}
\end{equation}
for $B\to K \nu\bar{\nu}$ and $B\to \pi \nu \bar{\nu}$ decays, respectively
\cite{Jeon:2006nq}.
The present exclusive mode gives more stringent bounds on the leptophobic
FCNC coupling compared with the inclusive $b\to s \nu\bar{\nu}$ decay
\cite{Leroux:2001fx}.

Recently, the Belle Collaboration has reported
upper limits on the production of the $K^\ast$ meson  with the missing
energy signal at the $B$ decay
where its BR is expected to be about 3 times larger than
that of the scalar meson production in the SM~\cite{:2006vg}.
It provides the constraint on the FCNC coupling
\begin{equation}
|U_{sb}^{Z^\prime}| \leq 0.66,
\end{equation}
which is larger than that in Eq.~(\ref{Ubound}).
At the super-$B$ factory, all four decay modes $B\to M\nu \bar{\nu}$
would be well measured and give more stringent bounds on the FCNC couplings.

The exclusive modes are much easier at the experimental detection than
the inclusive ones.
However, the exclusive modes have inevitable large theoretical
uncertainties from hadronic transition form factors.
In order to reduce hadronic uncertainties,
one can take ratios for  ${\cal B}(B\to M \nu \bar{\nu})$
to ${\cal B}(B\to M e \nu)$ for $M=\pi,\rho$ mesons\cite{Jeon:2006nq}.

\subsection{$B_s^0-\bar{B}_s^0$ Mixing
\label{sec4}}

The $Z^\prime$-exchanging $\Delta B = \Delta S = 2$
tree diagram contributes to the $B_s^0-\overline{B}_s^0$ mixing~\cite{Baek:2006bv}.
The mass difference $\Delta m_s$ of the mixing parameters then
read
\begin{eqnarray}
\Delta m_s = \Delta m_s^{\rm SM}
   \left|1 + R ~e^{2i \phi_{sb}^{Z^\prime}} \right|,
\end{eqnarray}
\begin{eqnarray}
R \equiv  \frac{2\sqrt{2} \pi^2}
     {G_F M_W^2 \left( V_{tb} V{_{ts}^\ast} \right)^2  S_0(x_t)}
     \frac{M_Z^2}{M_{Z^\prime}^2}
     \left|U_{sb}^{Z^\prime}\right|^2
     = 1.62 \times 10^3
       \left(\frac{700 ~\rm{GeV}}{M_{Z^\prime}}\right)^2
       \left|U_{sb}^{Z^\prime}\right|^2.
\end{eqnarray}

\begin{figure}
\begin{tabular}{cc}
\psfig{file=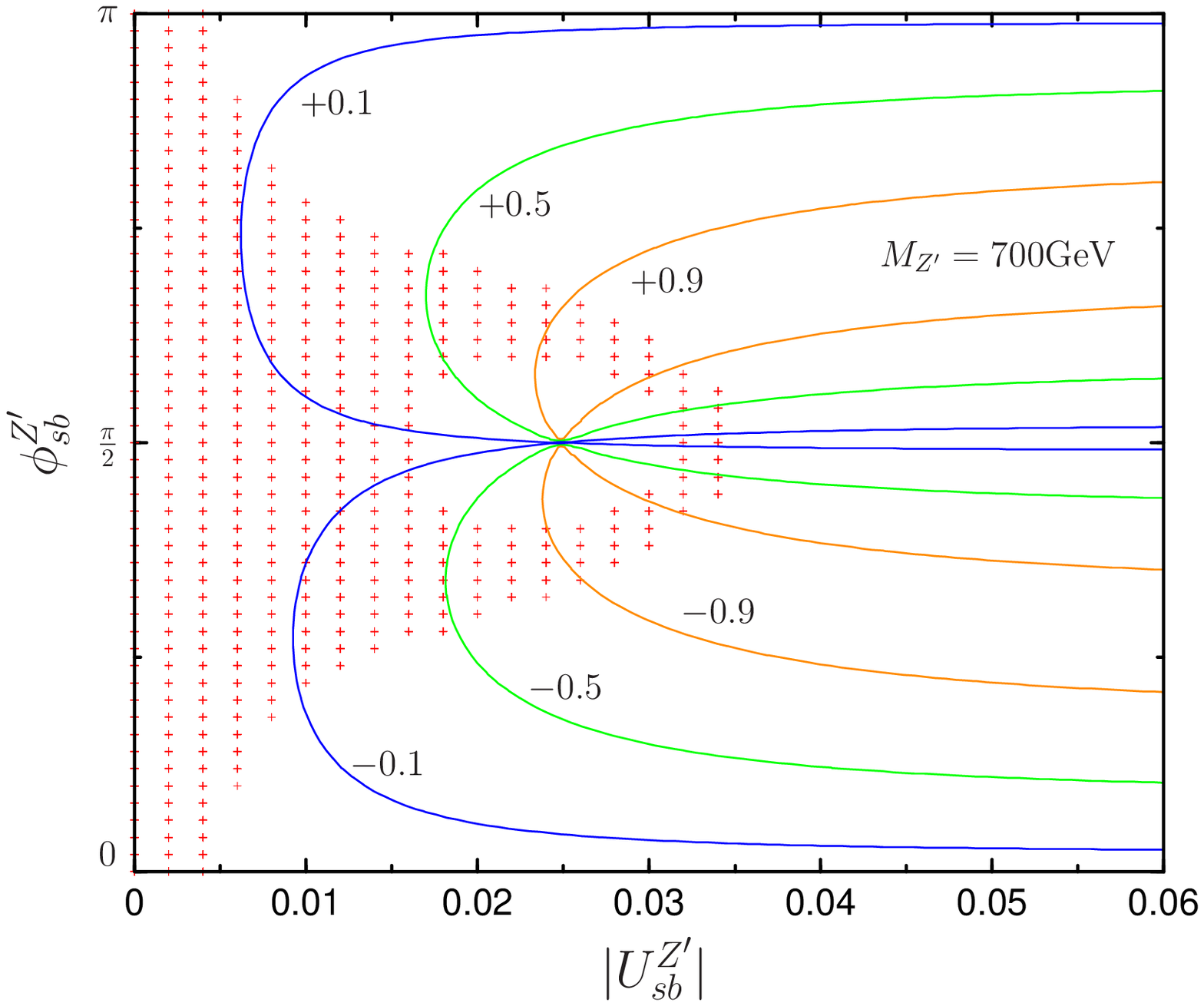,width=6cm}~~~&
\psfig{file=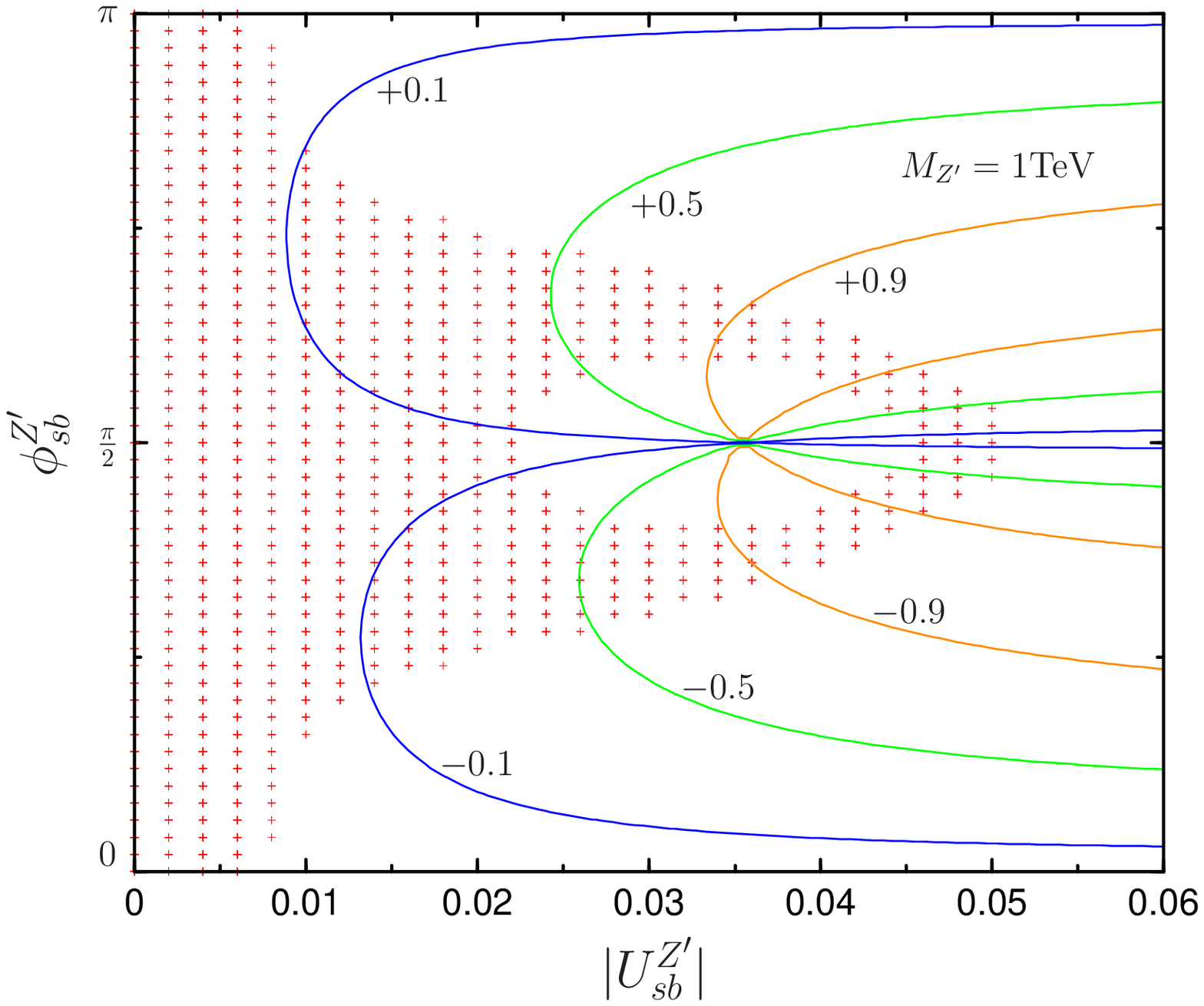,width=6cm}~~~\\[-0.5ex]
\textbf{(a)}&\textbf{(b)}
\end{tabular}
\vspace*{1pt}
\caption{ \label{fig3} The allowed region in
($|U_{sb}^{Z^\prime}|$,$\phi_{sb}^{Z^\prime}$) plane for
(a)~$M_{Z^\prime}=700$ GeV and (b)~$M_{Z^\prime}=1$ TeV~. We used
(HP+JL)QCD result  for the hadronic parameter.
Constant contour lines for the time dependent CP asymmetry
$S_{\psi\phi}$ in $B_s \to J/\psi~\phi$ are also shown. }
\end{figure}

In Figs.~\ref{fig3}, the allowed region in
($|U_{sb}^{Z^\prime}|$,$\phi_{sb}^{Z^\prime}$) plane is shown.
We obtain \begin{equation} |U_{sb}^{Z^\prime}|
\leq 0.0055 \qquad   \rm{for} ~M_{Z^\prime} = 700 ~ \rm{GeV},
\end{equation}
for $\phi_{sb}^{Z^\prime}=0$.
This bound is about two orders of magnitude stronger than
(\ref{Ubound}) obtained from exclusive semileptonic
$B \to M \nu \bar{\nu}$ decays.

The holes appear because they predict too small $\Delta m_s$.
For a given $M_{Z^\prime}$ we can see that large CP violating phase can enhance
the allowed coupling $|U_{sb}^{Z^\prime}|$ up to almost factor 10.
This shows the importance of the role played by CP violating phase
even in CP conserving observable such as  $\Delta m_s$.
As can be seen from Fig. 3(b),
irrespective of its phase $\phi_{sb}^{Z^\prime}$ value
\begin{equation}
|U_{sb}^{Z^\prime}| \leq 0.051 \qquad
                    \rm{for} ~M_{Z^\prime} = 1 ~ \rm{TeV}.
\end{equation}

The CP violating phase in $B_s^0 - \overline{B}_s^0$ mixing amplitude
can be measured at LHC in near future through the
time-dependent CP asymmetry in $B_s \to J/\psi~\phi$ decay
\begin{equation}
 \frac{ \Gamma \left(\overline{B}_s^0(t) \to J/\psi~ \phi \right)
       -\Gamma \left(B_s^0(t) \to J/\psi~ \phi \right)}
      { \Gamma \left(\overline{B}_s^0(t) \to J/\psi~ \phi \right)
       +\Gamma \left(B_s^0(t) \to J/\psi~ \phi \right)}
 \equiv S_{\psi\phi} \sin \left(\Delta m_s t\right).
\end{equation}
We note that although the final states are not CP-eigenstates, the time-dependent
analysis of the $B_s^0 \to J/\psi~ \phi$ angular distribution allows a clean
extraction of $S_{\psi\phi}$~\cite{angular}.
In the SM, $S_{\psi\phi}$ is predicted to be very small,
$S_{\psi\phi}^{\rm SM} =-\sin 2\beta_s =0.038 \pm 0.003$
$\left(\beta_s \equiv \arg \left[(V_{ts}^* V_{tb}) / (V_{cs}^*
V_{cb})\right]\right)$. If NP has an additional CP violating phase
$\phi_{sb}^{Z^\prime}$, however, the experimental value of
\begin{equation}
 S_{\psi\phi} = -\sin \left[ 2 \beta_s +
                            \arg \left(1 + R ~e^{2i \phi_{sb}^{Z^\prime}} \right)
                     \right]
\end{equation}
would be significantly different from the SM prediction.
Constant contour lines for $S_{\psi\phi}$ are also shown in Figs.~\ref{fig3}.
We can see that even with the strong constraint from the present $\Delta m_s$ observation,
large $S_{\psi\phi}$ are still allowed.

\section*{Acknowledgements}

CSK is supported in part by CHEP-SRC and in part
by the KRF Grant funded by the Korean Government (MOEHRD)
No. KRF-2005-070-C00030.

\end{document}